\documentclass[10pt,aps,prd,twocolumn,showpacs,nofootinbib]{revtex4-1}
\pdfoutput=1
\usepackage{graphicx}
\usepackage{dcolumn}
\usepackage{amsmath}
\usepackage{amssymb}
\usepackage{hyperref}

\begin{document}

\title{Resonantly-Produced 7 keV Sterile Neutrino Dark Matter Models \\ and
  the Properties of Milky Way Satellites}

\author{Kevork N.\ Abazajian}
\affiliation{Center for Cosmology, Department of Physics and Astronomy, University of California, Irvine, Irvine, California 92697 USA}

\date{March 4, 2014}

\begin{abstract}
Sterile neutrinos produced through a resonant Shi-Fuller mechanism are
arguably the simplest model for a dark matter interpretation origin of
the recent unidentified X-ray line seen toward a number of objects
harboring dark matter. Here, I calculate the exact parameters required
in this mechanism to produce the signal. The suppression of small
scale structure predicted by these models is consistent with Local
Group and high-$z$ galaxy count constraints. Very significantly, the
parameters necessary in these models to produce the full dark matter
density fulfill previously determined requirements to successfully
match the Milky Way Galaxy's total satellite abundance, the
satellites' radial distribution and their mass density profile, or
``too big to fail problem.'' I also discuss how further precision
determinations of the detailed properties of the candidate sterile
neutrino dark matter can probe the nature of the quark-hadron
transition, which takes place during the dark matter production.
\end{abstract}

 
\maketitle

\noindent {\it Introduction ---} There has been interest in dark
matter having ``warm'' properties for a significant amount of
time~\cite{Blumenthal:1982mv}. Interest in warm dark matter (WDM)
increased in proposed solutions to the ``missing satellites'' problem
of the Local Group of galaxies \cite{Bode:2000gq}. A single-particle
sterile neutrino addition to the standard model of particle physics
provides a minimalist extension that can be produced as a
WDM particle in negligible (standard) lepton number cosmologies
through non-resonant collision-dominated production via neutrino
oscillations in the early Universe in the Dodelson-Widrow mechanism
\cite{Dodelson:1993je}. This mechanism has a narrow range of parameters that
suffice the requirement of providing the observed dark matter
density. Resonant Mikheyev-Smirnov-Wolfenstein (MSW) mechanism production of
sterile neutrino dark matter was found to occur in cosmologies with a
small but significant lepton number through the Shi-Fuller mechanism,
and the sterile neutrinos could be produced with the proper abundance
with a wider range of parameters~\cite{Shi:1998km}. Both of these
production mechanisms were forecast to have a radiative decay mode
that could be detectable by X-ray telescopes as an unidentified line
in observations of X-ray clusters and in field galaxies
\cite{Abazajian:2001nj,Abazajian:2001vt}.

\begin{figure}[t!]
\includegraphics[width=3.43truein]{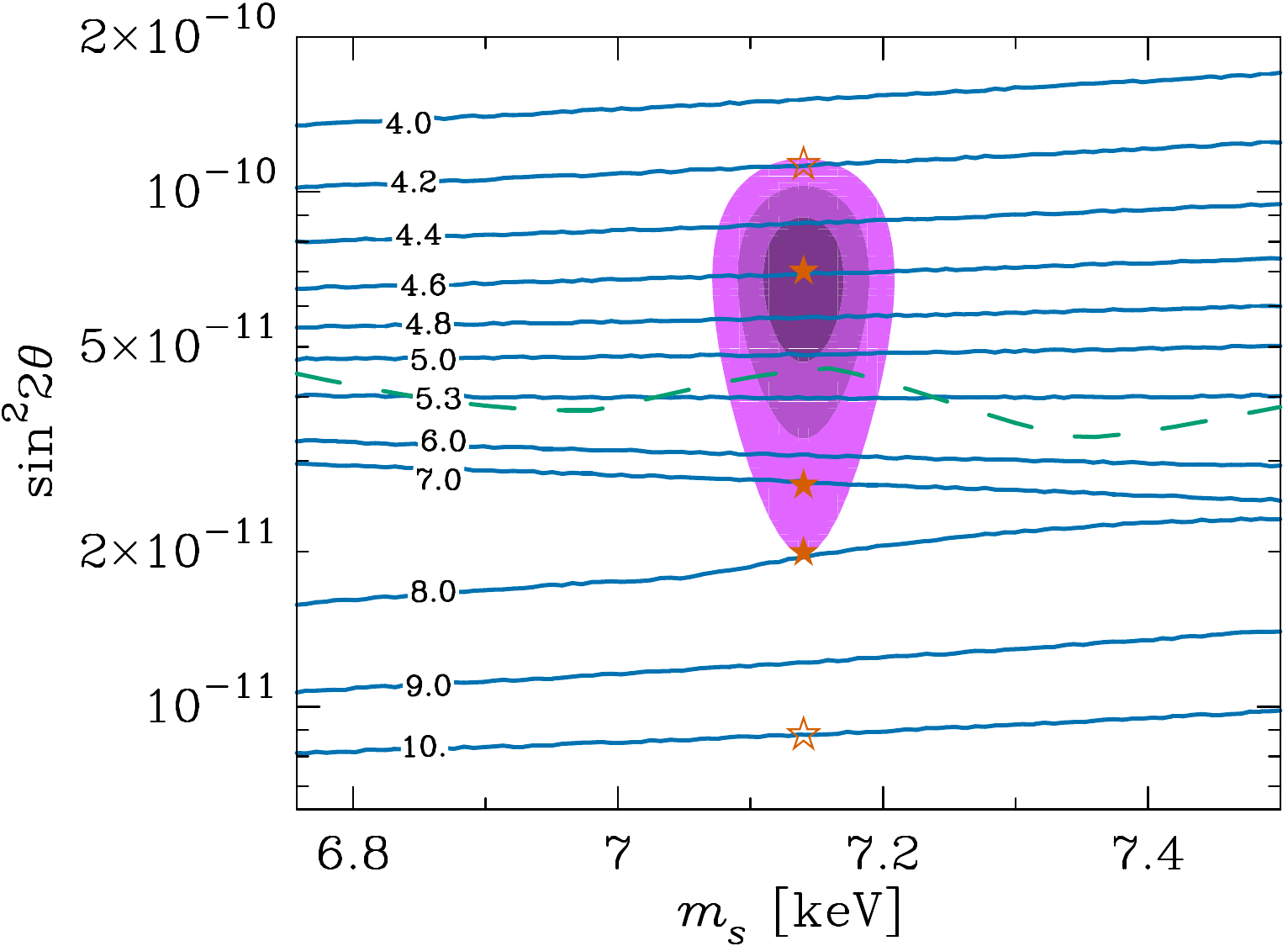}
\caption{\footnotesize This illustrates the parameter space for Shi-Fuller resonant
  production sterile neutrino models in the region of interest for
  producing the unidentified 3.57 keV X-ray line. The filled colored
  contours are the 1, 2 and 3$\sigma$ regions satisfying the
  best-determined unidentified line flux in the 6 Ms {\it XMM-Newton}
  73 stacked-cluster sample of Bulbul et
  al.~\cite{Bulbul:2014sua}. Systematic uncertainties on the flux and
  mixing angle are of order the 2$\sigma$ uncertainties. The blue,
  approximately horizontal contours are labeled by the lepton number
  $L_4$, in units of $10^{-4}$, needed to produce $\Omega_{\rm DM}
  h^2= 0.119$. The constraint from X-ray observations of M31 from
  Horiuchi et al.~\cite{Horiuchi:2013noa} are in dashed (green), with
  a notable upturn at the signal region. The five stars produce
  the phase space distributions shown in Fig.~\ref{distrib}, and the
  three solid stars produce the linear WDM power spectrum transfer
  functions in Fig.~\ref{transfer}. The contours change their
  orientation because the primary temperature of resonant production
  of the sterile neutrinos changes from prior to the quark-hadron
  transition to after it with increasing lepton numbers, for the case
  of the standard cross-over quark-hadron transition at $T_{\rm QCD} =
  170\rm\ MeV$~\cite{Abazajian:2002yz}. 
  \label{leptoncontour}}
\vskip-0.7 cm
\end{figure}

The Dodelson-Widrow model has recently been shown to be in significant
conflict with constraints from the Local Group, namely, the galaxies'
phase space densities, their subhalo abundance and X-ray observations
of the Andromeda galaxy (M31) by Horiuchi et
al.~\cite{Horiuchi:2013noa}, and is also constrained significantly by measures of
clustering in the Lyman-$\alpha$ (Ly-$\alpha$) forest
\cite{Seljak:2006qw,Viel:2013fqw}. However, lepton-number driven
production occurs for smaller neutrino mixing angles in universes with
lepton numbers significantly larger than the cosmological baryon
number, which may be generated by the Affleck-Dine mechanism
\cite{Affleck:1984fy} or via massive particle
decay~\cite{Laine:2008pg}. Recently, there have been reports of the
detection of an unidentified line in stacked observations of X-ray
clusters with the {\it XMM-Newton X-ray Space telescope} with both CCD
instruments aboard the telescope, and the Perseus cluster with the
{\it Chandra X-ray Space Telescope}~\cite{Bulbul:2014sua}.  An
independent group has indications of a consistent line in {\it
  XMM-Newton} observations of M31 and the Perseus
Cluster~\cite{Boyarsky:2014jta}. Interestingly, these detections lay
at the edge of the X-ray constraints from Horiuchi et al. (see
Fig.~\ref{leptoncontour}).

In this {\it Letter}, I calculate the details of the production and
perturbation evolution of resonantly-produced sterile neutrino models
that satisfy the signal for the best-determined parameters producing
the flux of the line. The mass and rate is that determined by the 6 Ms
of 73 stacked-cluster observations by the {\it XMM-Newton} MOS
detector sample of Bulbul et al.~\cite{Bulbul:2014sua}. First, I
calculate the combination of lepton numbers, particle mass and mixings
required to produce the signal, which are found to be consistent with
other constraints. In addition, I calculate the evolution of the
perturbations of sterile neutrino dark matter through the early
Universe that lead to cosmological structure formation from large
scale structure to sub-Milky-Way structure. I show that
some of the parameter space in the resulting models are consistent with the Local
Group constraints in Horiuchi et
al.~\cite{Horiuchi:2013noa}. Furthermore, the transfer functions in
the region produce a suppression
in power at small scales that was found previously to be preferred to
satisfy the properties of Milky Way satellites
\cite{Anderhalden:2012jc}, including the total satellite abundance,
their radial distribution and their mass profile
\cite*{BoylanKolchin:2011de,BoylanKolchin:2011dk,Lovell:2011rd}. The
last has been dubbed the ``too big to fail problem'' since the halos
in the models with the full cold dark matter (CDM) power produce halos with
densities that should not fail to produce Milky Way satellite
galaxies, but are, in turn, not seen.

\begin{figure}[t!]
\includegraphics[width=3.43truein]{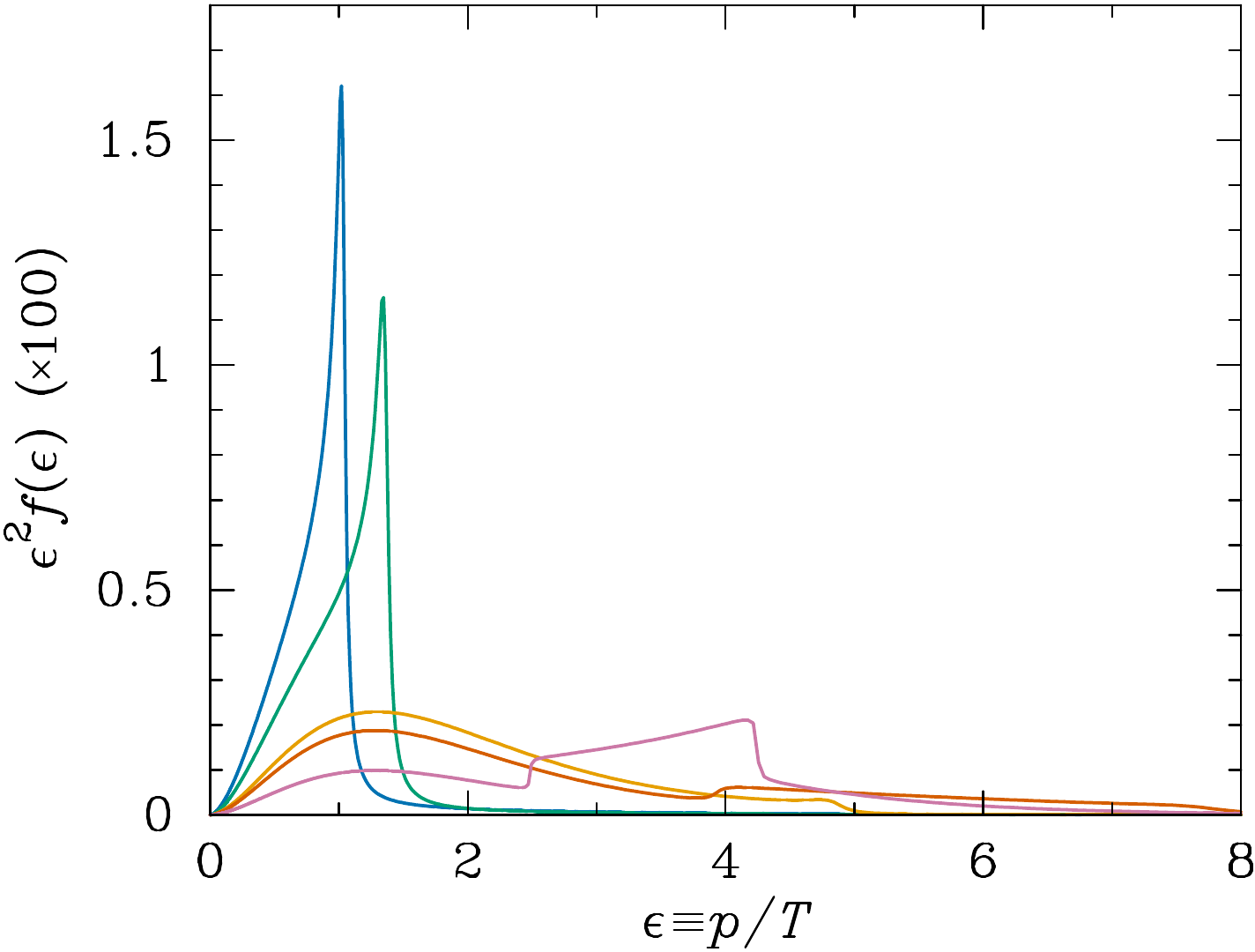}
\caption{\footnotesize Shown here are the distribution functions of the 7.14 keV
  models shown as stars in Fig.~\ref{leptoncontour}. The models with
  $L_4=4.2, 4.6, 7, 8, \text{and}\ 10$ have, respectively, increasing
  average $\langle p/T\rangle$, and therefore larger-scale cutoffs in
  the linear matter power spectrum for the fixed particle mass. The
  4.2 and 4.4 models have resonant production almost entirely prior to
  the quark-hadron transition, and therefore significantly ``colder''
  properties than the remaining models, whose step-function-like
  features are due the quasi-isotemperature evolution of the position
  of the resonance during the quark-hadron transition. All
  distributions are thermally cooler than the corresponding
  Dodelson-Widrow case, where $\langle p/T\rangle \approx 3.15$.
  \label{distrib}}
\vskip -0.7 cm
\end{figure}

\noindent{\it Production Calculations ---} To calculate the resonant
production of the sterile neutrino, I follow the detailed resonant
production calculations first performed in
Ref.~\cite{Abazajian:2001nj} and specified for properties of the
quark-hadron transition in Ref.~\cite{Abazajian:2002yz}.

The resonance in the production of the sterile neutrinos has the
momentum position of
\begin{eqnarray}
\label{eres}
\epsilon_{\rm res} &\approx& {\frac{\delta m^2}{
{\left( 8\sqrt{2}\zeta(3)/\pi^2\right)} G_{\rm F} T^4 {{L}} }}\\
&\approx& 3.65 {\left({\frac{\delta m^2}{(7\,{\rm
keV})^2}}\right)} {\left({\frac{{10}^{-3}}{{{L}}}}\right)}
{\left({\frac{170\,{\rm MeV}}{T}}\right)}^4 \nonumber\ ,
\end{eqnarray}
where $\epsilon_{\rm res} \equiv p/T|_{\rm res}$ is the position of the
resonance, $\delta m^2 \equiv m_2^2-m_1^2$, where $m_2$ is more
identified with the sterile neutrino. Here, $L \equiv
\left(n_{\nu_\alpha} - n_{\bar\nu_\alpha}\right)/{n_\gamma}$ is the
lepton number of the Universe prior to resonant production, relative
to the photon number $n_\gamma$. Since the lepton numbers of interest
are of order $10^{-4}$, we define $L_4 \equiv 10^4 L$. Here, the
calculation is done for the flavor $\alpha = \mu$, but the general
features of the calculation are independent of flavor. There are
subtleties with the effects of quantum-Zeno-effect damping in the full
quantum kinetic equations (QKEs) in the case
of resonance \cite{Boyanovsky:2006it}, but tests with
the full QKEs in the resonance find quasi-classical
quantum-Zeno treatment of production as adopted here is
appropriate~\cite{Kishimoto:2008ic}. Further tests of the production
in these models with the full QKEs is warranted, but beyond the scope
of this brief {\it Letter}.

As discussed in Ref.~\cite{Abazajian:2001nj}, as the Universe expands
and cools with time, for a given $\delta m^2$, the resonance will
sweep through the $\nu_\alpha$ energy distribution function from low
to high neutrino spectral parameter $\epsilon$. Before peak
production, the sweep rate is $d\epsilon/dt \approx 4
\epsilon H \left( 1-\dot L/4 H L\right)$, where ${\dot{{L}}}$ is the
time rate of change of the lepton number resulting from neutrino
flavor conversion, and $H$ is the expansion of the Universe.

The dominant effect on production is the value of the lepton number,
which in turn sets the required $\sin^2 2\theta$ to get the
cosmologically observed $\Omega_{\rm DM} h^2$. Because of this
dependence, and since the production is largely independent of the
sterile neutrino particle mass, we fix $m_s = 7.14\rm\ keV$, and
explore how production changes with different values of 
$L_4 = 4.2, 4.6, 7, 8, \text{and}\ 10$, shown as stars in Fig.~\ref{leptoncontour}.

As discussed in Ref.~\cite{Abazajian:2001nj}, since the expansion rate
scales as $H\sim T^2$, the prospects for adiabaticity (efficiency) of
the resonance are better at lower temperatures and later epochs in the
early Universe, all other parameters being the same, up until the
lepton number is depleted, and conversion ceases.  This produces the
increasing peak in the distribution function for $L_4 =
4.2\ \text{and}\ 4.6$ models. For larger lepton numbers, the resonance
through the momentum distributions is at lower temperatures, partially
before and partially after the quark-hadron transition, which is
readily seen in the scaling of Eq.~\eqref{eres}. The
quasi-isotemperature evolution of the Universe during and after the
quark hadron transition due to the heating of plasma with quark,
massive hadron and pion disappearance at this time produces the
step-function feature in the distribution functions seen in the $L_4 =
7, 8, \text{and}\ 10$ cases, as shown in Fig.~\ref{distrib}.

\begin{figure}[t!]
\includegraphics[width=3.43truein]{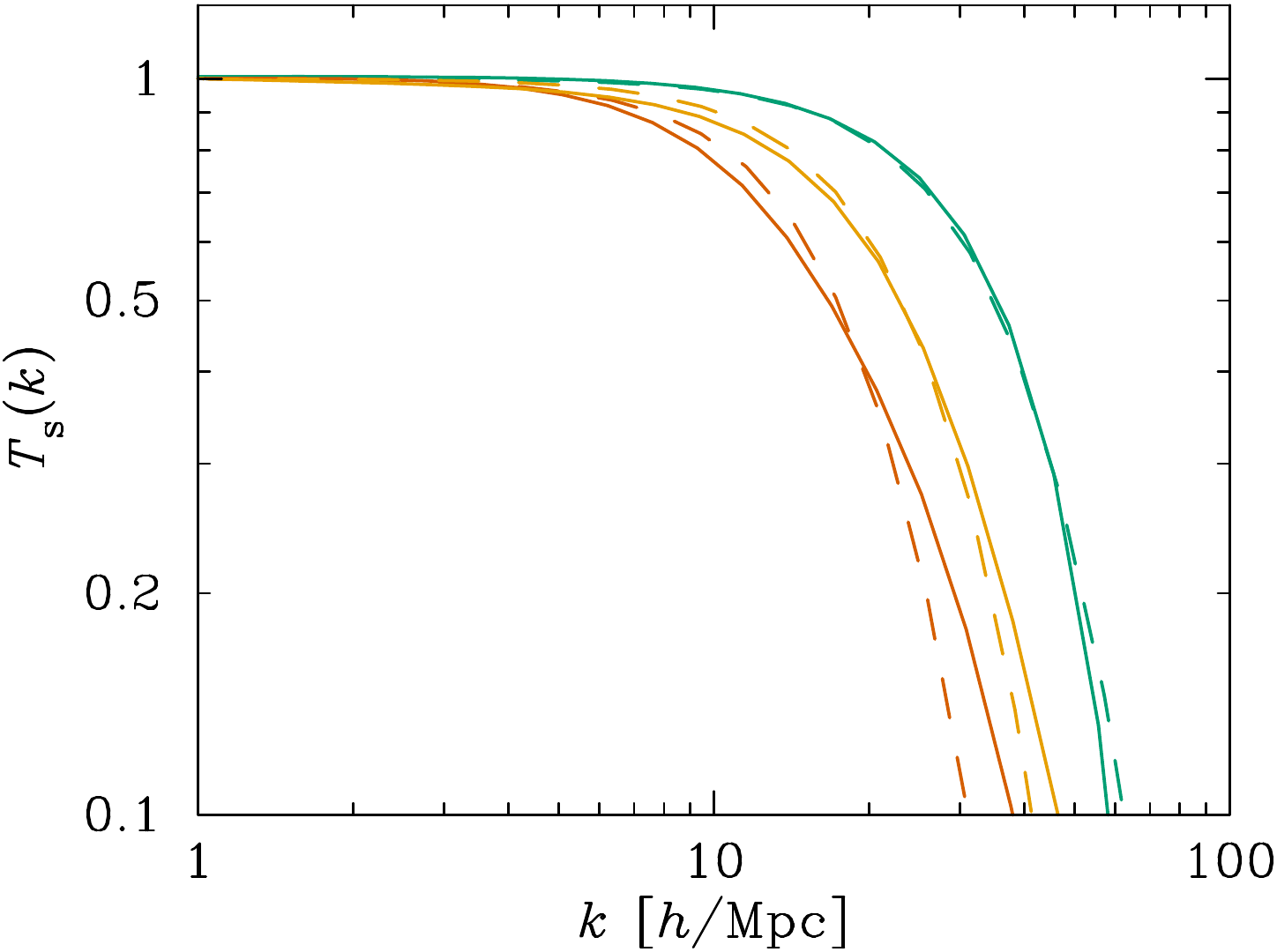}
\caption{\footnotesize Shown here are the WDM sterile neutrino
  transfer functions $T_s(k)\equiv \sqrt{P_{\rm WDM}(k)/P_{\rm
      CDM}(k)}$ for the three solid stars in Fig.~\ref{leptoncontour},
  as solid lines, with $L_4 = 8, 7\ \text{and}\ 4.6$ having increasing
  $k$ cutoff scale, respectively. Colors correspond to the
  distibutions in Fig.~\ref{distrib}. The dashed lines are the thermal
  WDM transfer functions that best fit these cases, which have a
  thermal WDM particle mass of $m_{\rm thermal} = 1.6,
  2.0\ \text{and}\ 2.9\ \rm keV$, respectively.
  \label{transfer}}
\vskip -0.7 cm
\end{figure}

\noindent{\it Linear Cosmological Structure Inferred Properties \&
  Constraints ---} The resonant production mechanism was known since
its introduction to produce thermally ``cooler'' distributions than in
the Dodelson-Widrow mechanism~\cite{Shi:1998km}. To accurately
determine the nature of the matter perturbations arising in these
models, I calculate the evolution of the sterile neutrino dark matter
perturbations for the three representative cases $L_4 = 4.6, 7,
\text{and}\ 8$, as in Ref.~\cite{Abazajian:2005gj}, using the
numerical Boltzmann cosmological perturbation evolution solver
CAMB~\cite{Lewis:1999bs}. The resulting suppression of perturbations
relative to CDM, $T_s(k)^2 = P_{\rm WDM}(k)/P_{\rm CDM}(k)$, is shown in
Fig.~\ref{transfer}. The exact shape of the transfer functions are
different from the case of thermal WDM~\cite{Bode:2000gq}, but can be
fit by thermal WDM transfer functions, shown as dashed lines in
Fig.~\ref{transfer}. The representative cases of $L_4 = 4.6, 7,
\text{and}\ 8$ are best fit by thermal WDM particle masses of 1.6, 2.0
and 2.9 keV, respectively.

The most stringent claimed constraint on WDM is from observations and
modeling of the Ly-$\alpha$ forest seen in spectra toward distant
quasars. However, the Ly-$\alpha$ forest is a complicated tool for
inferring the linear matter power spectrum, requiring disentangling
the effects of pressure support and thermal broadening of the
Ly-$\alpha$ forest features from the effects of dark matter
perturbation suppression from WDM. In addition, modeling the
dependence on the physics of the neutral gas requires assumptions of
the thermal history of the intergalactic medium and its ionizing
background. These are done as parameterized fitting functions. Many of
the limitations of the Ly-$\alpha$ forest on constraints of the
primordial power spectrum are discussed in Abazajian et
al. \cite{Abazajian:2011dt}. The Ly-$\alpha$ forest modeling of the
inter-galactic medium and its numerical methods have
considerably improved in recent years, and can provide
stringent constraints which should be further studied regarding their
robustness. Recent quoted limits are at thermal particle masses of
$m_{\rm thermal} > 3.3\rm\ keV$ (95\% CL) \cite{Viel:2013fqw}.

Horiuchi et al.~\cite{Horiuchi:2013noa} studied in detail the phase
space constraints from Local Group dwarfs on Dodelson-Widrow sterile
neutrino dark matter models. The warmest resonant dark matter model
considered here, $L_4 = 10$ has a phase space density $Q_{\rm max}
\approx 398$, well above the minimum values inferred by Local
Group dwarfs, e.g., Segue 1's requirement of $Q_{\rm sim} \approx
30$~\cite{Horiuchi:2013noa}.  Horiuchi et al.~\cite{Horiuchi:2013noa}
and Polisensky \& Ricotti \cite{Polisensky:2010rw} studied the
constraints from the minimal subhalo number counts required to produce
the observed dwarf galaxy populations of the Local Group. The
constraint from subhalo counts in Horiuchi et
al.~\cite{Horiuchi:2013noa} on Dodelson-Widrow sterile neutrinos
corresponds to a thermal WDM particle mass of $m_{\rm thermal} >
1.7\rm\ keV$ (95\% CL). In addition, high-$z$ galaxy counts exclude
thermal WDM particle masses below $m_{\rm thermal} = 1.3\rm\ keV$ at least at
2.2$\sigma$~\cite{Schultz:2014eia}. The high-$z$ and Local Group
subhalo constraints are in conflict with the warmest of the three
resonant models for which we demonstrate the linear matter
transformation function, $L_4 = 8$. The $L_4 = 4.6\ \&\ 7$ cases survive
these structure formation constraints.

As shown in Fig.~\ref{leptoncontour}, the central $L_4 = 4.6$ case is
in tension with the constraints from stacked X-ray observations of
M31~\cite{Horiuchi:2013noa}. The $L_4 = 7$ case is consistent with all
Local Group structure formation and X-ray constraints, with a transfer
function that matches that of thermal WDM of $m_{\rm thermal} = 2.02\ \rm keV$.

Perhaps most interestingly, the case of thermal WDM of $m_{\rm thermal} =
2\rm\ keV$ is the cutoff scale inferred as a solution to the Milky Way
satellite's total satellite abundance, the satellites' radial
distribution and their mass density profile, or ``too big to fail
problem,'' first discussed in Lovell et al.~\cite{Lovell:2011rd} and
explored in detail in Anderhalden et al.~\cite{Anderhalden:2012jc},
Polisensky \& Ricotti \cite{Polisensky:2013ppa} \& Kennedy et
al.~\cite{Kennedy:2013uta}. Thermal WDM particle masses of $m_{\rm
  thermal} > 2\rm\ keV$ are thought to not have significant difference
with CDM in subhalo kinematics (i.e., inner densities)
\cite{Schneider:2013wwa}. These issues are also reviewed in detail in
Ref.~\cite{Weinberg:2013aya}. Overall, the range of thermal $m_{\rm
  thermal} = 1.7 - 2.0\rm\ keV$ is a ``sweet spot'' that may address the
controversies in structure formation at small scales. Remarkably, as
shown here, the cutoff scale of these WDM models is consistent with
the preferred $L_4 \approx 7$ resonant sterile neutrino dark matter
models of the unidentified X-ray line.

\noindent{\it Conclusions ---} Here I have presented the calculation
of resonantly-driven MSW mechanism production of sterile neutrino dark
matter that is in the parameter space of interest for the sterile
neutrino dark matter decay interpretations of the recently reported
unidentified X-ray line in X-ray clusters and M31, with the ansatz
that sterile neutrinos are all of the dark matter. I have also
presented the linear perturbation evolution and resulting transfer
functions of the cool-to-warm resonant models that are consistent the
X-ray signal. Significantly, if the signal was slightly larger in
$\sin^2 2\theta$ or $m_s$, then the effective thermal WDM mass would
have been very cold and unconstrained, but commensurately less of
interest. If $\sin^2 2\theta$ or $m_s$ was slightly smaller, then the
effective thermal WDM mass would have been very warm and constrained
by several constraints that exist on WDM. However, the cutoff of the
perturbation dark matter power spectrum in these models straddles
exactly the region of interest to alleviate structure formation
controversies at small scales, corresponding to thermal WDM particle
masses of 1.7 to 2.9 keV. The lepton number model with $L_4=7$ is
consistent with producing the 3.57 keV X-ray line, has a linear
structure cutoff that matches that of thermal WDM particle masses of 2
keV and is consistent with constraints from the Local Group, high-$z$
galaxy counts as well as other X-ray constraints. Further tests of
whether the Ly-$\alpha$ forest constraints are robust in this
parameter space are certainly warranted.

Confirmation or exclusion of the X-ray line as originating from dark
matter is certainly of great interest. This has been discussed in the
literature, and test of the dark matter origin may be done via (1)
X-ray observations of more Local Group galaxies
\cite{Boyarsky:2006fg}, (2) energy resolution of the line to have dark
matter velocity broadening instead of thermal broadening of a plasma
line with the upcoming {\it Astro-H} mission~\cite{Bulbul:2014sua}, (3)
accurate determination of the flux profile of the line which could
distinguish it from the plasma line's flux profile as well as dark
matter models that involve two-body processes and therefore a
density-squared profile \cite{Finkbeiner:2014sja}. Lastly, to connect
the X-ray line feature to the neutrino sector, it is arguable that it
should be confirmed, e.g., by kink searches in nuclear $\beta$-decay spectra
\cite{deVega:2011xh}.

Thinking further ahead, as discussed in Ref.~\cite{Abazajian:2002yz}
and discussed above, if the 7 keV sterile neutrino is determined to be
the dark matter, and the properties of the sterile neutrino required to
produce the signal are sufficiently well-determined, then one could
hope to eventually test the nature of the quark-hadron transition
which occurs during the resonant sterile neutrino dark matter
production, since its properties affect the production.

It remains to be seen if the consistency between the unidentified
X-ray line with small scale structure formation considerations is a
mere coincidence or the emergence of a convergence of astrophysical
methods of determining the nature of dark matter.

\noindent {\it Acknowledgments --- } I would like to thank James
Bullock, George Fuller, Shunsaku Horiuchi, and Manoj Kaplinghat for
useful discussions. I would particularly like to thank Shunsaku
Horiuchi for providing calculations of the phase space densities of the
models. KNA is supported by NSF CAREER Grant No. PHY-11-59224.

\bibliography{/Users/aba/Dropbox/master.bib}

\end{document}